\documentclass[fleqn,usenatbib,useAMS]{mnras}

\usepackage[T1]{fontenc}
\usepackage{ae,aecompl}

\usepackage{graphicx}	
\usepackage{amsmath}	
\usepackage{amssymb}	
\usepackage{mathptmx}
\usepackage{txfonts}
\usepackage{hyperref}
\usepackage{breakurl}

\title[Magnetism in cool, white dwarfs]{Evidence of enhanced magnetism in cool, 
polluted white dwarfs}

\author[A. Kawka et al.]{Adela Kawka$^{1}$\thanks{E-mail: adela.kawka@curtin.edu.au (AK)}\thanks{Visitor at the Mathematical Sciences Institute, The Australian National University, Canberra, ACT 0200, Australia}, 
St\'ephane Vennes$^{2}$,
Lilia Ferrario$^2$ and Ernst Paunzen$^3$ \\
$^{1}$ International Centre for Radio Astronomy Research - Curtin University, 
GPO Box U1987, Perth, WA 6845, Australia\\
$^{2}$ Mathematical Sciences Institute, The Australian National University,
Canberra, ACT 0200, Australia\\
$^{3}$ Department of Theoretical Physics and Astrophysics, Masaryk University, 
Kotl\'a\v{r}sk\'a 2, CZ-611 37, Czech Republic\\
}

\date{Accepted XXX. Received YYY; in original form ZZZ}

\pubyear{2017}

\begin{document}
\label{firstpage}
\pagerange{\pageref{firstpage}--\pageref{lastpage}}
\maketitle

\begin{abstract}
We report the discovery of a new, polluted, magnetic white dwarf in the Luyten 
survey of high-proper motion stars. High-dispersion spectra of NLTT~7547 reveal 
a complex heavy element line spectrum in a cool ($\approx 5\,200$~K) 
hydrogen-dominated atmosphere showing the effect of a surface averaged  
field of 163 kG, consistent with a 240 kG centred dipole,
although the actual field structure remains uncertain.
The abundance pattern shows the effect of accreted material with a distinct 
magnesium-rich flavour. Combined with earlier identifications, this discovery 
supports a correlation between the incidence of magnetism in cool 
white dwarfs and their contamination by heavy elements.
\end{abstract}


\section{Introduction}

Magnetic fields are observed in white dwarfs across the whole evolutionary
sequence with strengths varying from a few hundred MG down to a few 
kG \citep[see][for a review of magnetic white dwarfs]{fer2015}. The incidence
of magnetism within the white dwarf population remains uncertain. In 
magnitude limited surveys \citep[e.g.,][]{sch1995,kep2013} the incidence of
magnetism is $\approx 5\%$. In volume limited surveys of white dwarfs 
\citep[e.g.,][]{kaw2007} the incidence is estimated to be much higher (10 - 20 \%). 
\citet{lan2012} and \citet{bag2018b} conducted a survey of bright white dwarfs 
to search for weak magnetic fields and found that the incidence of fields in
the $\sim 1 - 10$~kG range is about 6$\pm$4\% which is higher than the 
1 - 2\% per decade incidence \citep{kaw2007} for stronger magnetic fields 
(100 kG - 500 MG). 

\citet{lie1979} argued that magnetism is more frequent among cooler, hence older
white dwarfs, than in their hotter and younger counterparts. The significance of
their conclusions was low due to their small sample of 7 magnetic white dwarfs
out of a sample of 448 white dwarfs. \citet{val1999} re-examined the 
correlation between incidence of magnetism and effective temperature using a 
larger sample of 49 magnetic white dwarfs out of a total of 1225 white dwarfs 
with reliable temperature estimates and concluded that the fraction of magnetic white 
dwarfs is indeed higher among cooler stars than among their hotter counterparts supporting 
the conclusions of \citet{lie1979}. On the other hand a 
recent study of magnetic field evolution along the white dwarf cooling sequence 
was conducted by \citet{fer2015} who plotted the magnetic field strength versus 
the effective temperature for a sample of over 600 white dwarfs. This plot did 
not show any apparent correlation between the magnetic field strength and 
effective temperature which led them to conclude that there is 
no field evolution along the white dwarf cooling track. The apparent lack of a
correlation between the field incidence and the effective temperature proposed
by \citet{fer2015} may be an artifact of their Sloan-dominated sample, where Sloan
spectra are low resolution and many have low signal-to-noise and therefore stronger magnetic fields
are more likely to be detected. 
Whether or not field decay takes place remains an open question. However, there is
clear evidence that the incidence of magnetism varies with spectral class and even
possibly with age. 

About a quarter of white dwarfs 
have atmospheres that are polluted by elements heavier than helium 
\citep{zuc2003,zuc2010}. The origin of these heavy elements has been shown 
to be in accreted planetary or asteroidal debris 
\citep[e.g.,][]{deb2002,kil2006,jur2007,jur2008}. \citet{kaw2014} have shown 
that the incidence of magnetism is higher in cool, hydrogen-rich polluted
(DAZ) white dwarfs. Similarly, \citet{hol2015} noted a higher incidence in  
cool, helium-rich, polluted (DZ) white dwarfs.  The hot DQ class of white 
dwarfs also shows an enhanced magnetic field incidence (with surface averaged 
fields ranging from $\sim 0.1$ up to $2.1$~ MG) that is close to 70\% of 
these objects \citep{duf2013}.

Field geometry may provide clues to the origin of magnetic fields in white 
dwarf stars. \citet{gar2012} predict complex geometries for magnetic white 
dwarfs formed via a dynamo event such as a merger, while descendents of Ap and 
Bp stars may be more likely to have simpler dipolar fields \citep{bra2004}. 
Field geometry is usually assumed to resemble a dipole or an offset dipole.
However, when more detailed spectroscopic and spectropolarimetric observations
are possible, these reveal that the magnetic field geometry
is more complex. Using Zeeman tomography, \citet{euc2005} 
showed that HE~1045$-$0908, which has an average surface field
strength of 16~MG, has a dominant quadrupolar field with contributions
from dipolar and octupolar field geometries. The field geometry of some
white dwarfs such as WD1953-011 can be modelled with a weak dipolar field
($B_d = 178\pm30$ kG) and a stronger field of $515\pm7$ kG localized on to a 
spot \citep{max2000,val2008}. The hot hydrogen-rich white dwarf 
EUVE~J0317-85.5 also appears to have an underlying field of $\sim 185$ MG 
with a high field spot of $\sim 425$ MG \citep{ven2003}. More recently,
\citet{lan2017} reported on the field topology of two magnetic white dwarfs
with fields less than 200~kG, with one star (WD~2047$+$372) having a field
topology corresponding to a simple dipole and the other (WD~2359$-$434)
having a more complex topology with an underlying dipolar field and an
overlaying quadrupolar field.

Hydrogen-rich white dwarfs develop a convection zone from effective temperatures
below about 14\,000~K, while helium-rich white dwarfs develop a convection 
zone below temperatures of about 28\,000~K. The suppression of convective 
motion by magnetic fields in white dwarf atmospheres was first investigated by 
\citet{dan1975} who demonstrated that the evolutionary cooling rate of white 
dwarfs depends on the magnetic field strength. Convective energy transfer was
also suppressed in the model atmospheres calculated by \citet{wic1986} when 
they were investigating the effect of magnetic blanketing in white dwarfs. 
The hypothesis that magnetic fields suppress convection in white dwarf 
atmospheres was recently revisited by \citet{val2014} who found that a 
slow down in white dwarf cooling explains the higher incidence of magnetism
among cool white dwarfs \citep{lie1979}. Using radiation 
magnetohydrodynamic simulations \citet{tre2015} proposed that convection 
is inhibited throughout the atmosphere by magnetic fields as low as 50~kG, 
but that the cooling rate will not be affected until the convection zone 
couples with the degenerate core. For higher masses, e.g., at $1.0\ M_\odot$ 
this occurs at $T_{\rm eff} \approx 6000$ K, for lower masses the coupling
occurs at lower temperatures. 

Our survey of high proper motion white dwarfs \citep{kaw2006,kaw2012a} has 
revealed several magnetic white dwarfs with effective temperatures below 
6000~K, including NLTT~10480 \citep{kaw2011} and NLTT~43806 \citep{zuc2011}.
NLTT~7547 was identified as a cool, polluted white dwarf by 
\citet{kaw2012a}. In this paper we show that NLTT~7547 also harbours a magnetic
field and contributes to the growing sample of magnetic cool DAZ white dwarfs.
In section \ref{sect_obs} we present the spectroscopy and photometry of
NLTT~7547 and in section \ref{sect_models} the analysis of our results, 
followed by a discussion in section 4 and a summary in section 5.

\section{Observations}\label{sect_obs}

The following section describes all available observations of NLTT~7547
and WD~0141-675. 
Table~\ref{tab_phot} lists the astrometric and photometric measurements.

\subsection{Spectroscopy}

We first observed NLTT~7547 with the FORS1 spectrograph attached to the Very 
Large Telescope (VLT) of the European Southern Observatory (ESO) at Paranal. 
We used the 600 lines mm$^{-1}$ grism (Grism 600B) centred at 4650 \AA\ and 
set the slit width to 1 arcsec providing a resolution of $\sim 6$ \AA. The 
observations were obtained on UT 2007 December 13 and conducted in the 
spectropolarimetric mode where we first obtained an exposure with the 
retarder plate rotated to $-45^\circ$ immediately followed by an exposure 
with the Wollaston prism rotated to $+45^\circ$. These datasets were already 
published as part of a spectropolarimetric survey in \citet{kaw2012a}.

The FORS1 spectra revealed NLTT~7547 to be a cool DAZ white dwarf, and 
therefore, we obtained a second set of observations with the X-shooter 
spectrograph \citep{ver2011} attached to the VLT at ESO. We obtained six
sets of spectra between UT 2015 August 24 and September 19 where we set the 
slit-width to 0.5, 0.9 and 0.6 arcsec for the UVB, VIS and NIR arms,
respectively. The exposure times for each visit were 2940, 3000 and five 
$\times$ 600 sec for UVB, VIS and NIR arms respectively. This setup provided a 
resolving power of 9760, 7410 and 8040 for the UVB, VIS and NIR arms, 
respectively.

We also retrieved archival spectrum of the newly identified cool DAZ 
WD~0141-675, which will contribute to our study of incidence of magnetism in 
cool DAZ white dwarfs. 
The spectrum was obtained with the X-shooter spectrograph on UT 2017 July 27.
The slit-width was set to 1.6, 1.5 and 1.2 arcsec for the UVB, VIS, and NIR,
respectively, with a nominal resolution of $\sim 3000$. However the spectra 
were taken with an average seeing of $\sim0.5$~arcsec at 5000 \AA\  and 
therefore we will assume a resolution of $R\approx5000$. The exposure times 
were $4\times390$, $4\times 300$ and $4\times220$ sec for UVB, VIS and NIR 
arms respectively.

\subsection{Photometry}

We collected photometric measurements from various surveys. The Sloan Digital 
Sky Survey (SDSS) provided optical photometry in $u$,$g$,$r$,$i$ and $z$ bands 
\citep{alb2017} and the Two Micron All Sky Survey (2MASS) provided
infrared photometry in $J$, $H$ and $K$ bands \citep{skr2006}. NLTT~7547 was 
not detected by the Galaxy Evolution Explorer (GALEX).

Finally, we obtained photometric time series of NLTT~7547 in $R$ using the 1.54~m 
telescope at La Silla. The series were obtained on UT 2017 October 18 and 
UT 2018 January 5. The first set of observations covered 404 minutes and 
the second set covered 205 minutes. The exposure times were 60 seconds for 
all observations at a cadence of one exposure every 78 seconds. Two comparison stars which were checked for variability
were used to conduct differential photometry of NLTT~7547.

\begin{table}
\centering
\caption{Astrometry and photometry}
\label{tab_phot}
\begin{tabular}{lcc} 
\hline
Parameter & Measurement & Reference \\
\hline
RA (J2000)                   & 02 17 19.63       & 1 \\
Dec (J2000)                  & $-$06 56 28.86    & 1 \\
$\mu_\alpha \cos{\delta}$ (\arcsec yr$^{-1}$) & $0.3907\pm0.0055$ & 1 \\
$\mu_\delta$ (\arcsec yr$^{-1}$) & $0.0143\pm0.0055$ & 1 \\
$\mu_\alpha \cos{\delta}$ (\arcsec yr$^{-1}$) & $0.4225\pm0.0003$ & 2 \\
$\mu_\delta$ (\arcsec yr$^{-1}$) & $0.0179\pm0.0002$ & 2 \\
$\pi$ (mas)                  & $22.553\pm0.162$  & 2 \\
$u$                          & $19.679\pm0.037$  & 3 \\
$g$                          & $18.318\pm0.007$  & 3 \\
$r$                          & $17.747\pm0.006$  & 3 \\
$i$                          & $17.528\pm0.007$  & 3 \\
$z$                          & $17.454\pm0.014$  & 3 \\
$G$                          & $17.800\pm0.002$  & 2 \\
$b_p$                        & $18.258\pm0.032$  & 2 \\
$r_p$                        & $17.218\pm0.032$  & 2 \\
2MASS $J$                    & $16.636\pm0.133$  & 4 \\
2MASS $H$                    & $16.182\pm0.216$  & 4 \\
2MASS $K$                    & $15.674\pm0.251$  & 4 \\
\hline
\end{tabular}\\
References: (1) \citet{sal2003}; (2) \citet{bro2018}; (3) \citet{alb2017};
(4) \citet{skr2006}
\end{table}

\section{Analysis}\label{sect_models}

In this section we describe the model atmospheres that we use to interpret the
optical spectra and thus derive the stellar parameters, the magnetic field 
strength and structure, and the abundance pattern. We also searched for 
possible spectroscopic and photometric variations.

\subsection{Models}

The model atmospheres assume convective equilibrium and line blanketing
by dominant elements \citep{kaw2012b}. Convection may be suppressed as needed.
We introduced in the spectral synthesis a dipolar field distribution following 
\citet{mar1984} and \citet{ach1989}. The dipole of strength $|B_p|$ may be 
offset along the polar axis by a fraction of the radius $a_z$ and inclined with 
respect to the viewer at an angle $i$. The visible surface is discretized in 
450 elements with 30 individual latitudes between 0 and 180$^\circ$ and, 
assuming field symmetry about the meridian, 15 longitudes between 0 and 
90$^\circ$ each counting for 2 elements. The local spectra were computed for 
each element. The surface-integrated spectrum is the sum of individual spectra, 
weighted by a linear limb-darkening coefficient.

\begin{figure}
\includegraphics[viewport=5 5 530 530,clip,width=0.9\columnwidth]{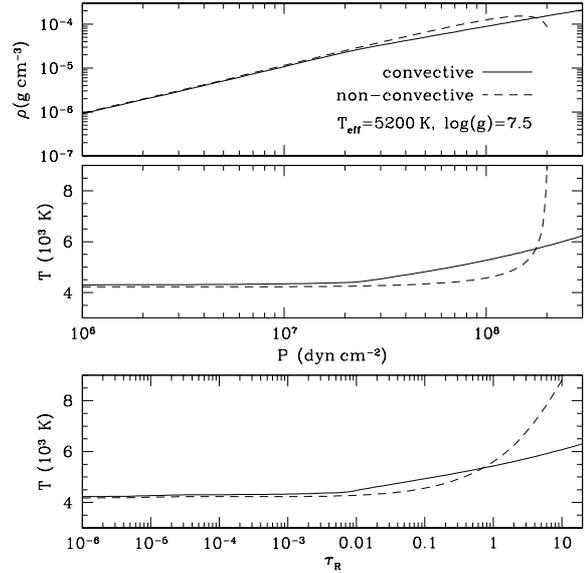}
\caption{A comparison of the structure of a model atmosphere with convective 
energy transport (full lines) and without (dashed lines). The temperature  
structure of the model at 5\,200\,K ($\log{g}=7.5$) is plotted as a function 
of optical depth (lower panel) and pressure (middle panel). The corresponding 
density structure is plotted as a function of pressure (upper panel).}
\label{fig_struct}
\end{figure}

Using convective models as an input we gradually suppressed the calculated 
convective flux in the models until it fell below 0.1\% of the total flux 
hence generating radiative atmospheres. In the process, the relatively 
shallow convective temperature gradient steepened sharply in order to 
sustain the increasing radiative outflow. The effect is evident when plotting 
the temperature as a function of Rosseland optical depths 
(Fig.~\ref{fig_struct}, lower panel). However, the resulting steep $dT/dP$ 
gradient in the non-convective model depicted in Fig~\ref{fig_struct} 
(middle panel) required adjustment to the depth discretization by concentrating 
the distribution where $dT/dP$ steepens. Fig~\ref{fig_struct} (upper panel) 
shows the corresponding density distribution. The non-convective models show 
a downturn in the density toward higher pressure which would give rise to 
a Rayleigh-Taylor instability. The same effect was noted in all models where 
convection was suppressed and further work should help elucidate the effect 
of these instabilities. We expect that the resulting mass motion should be 
accompanied by energy transfer and reduce the steepness of the temperature 
gradient.

The hydrogen Balmer spectra, from H$\alpha$ to H$\epsilon$ were computed using 
line strengths and Zeeman shifts  from \citet{sch2014}. We computed detailed 
heavy element line profiles following procedures described in \citet{kaw2011} 
but updated using quadratic Zeeman line splitting formulae from \citet{lan2004}.
The updated Zeeman patterns agree with earlier calculations for \ion{Ca}{ii} 
H\&K lines which employed the tabulation of \citet{kem1975}.

Given the low effective temperature of NLTT~7547, we tested whether the
population of molecules, such as NaH, MgH, AlH, CaH, and FeH, is  
high enough to affect the abundance measurements of the atomic species. We
conducted the tests in both convective and non-convective atmospheres. In our 
calculations we have used the partition functions of \citet{sau1984}, 
\citet{wen2010} and \citet{bar2016} with the FeH dissociation energy of 
1.69 eV \citep{che2017}. We found that abundance measurements can differ
by up to 10\% by including either MgH, AlH or FeH and about 1\% by including
NaH or CaH. We inspected the NIR spectrum and did not detect FeH in the 999 to
1000 nm range \citep[see, e.g.,][]{wen2010} although a spectral synthesis of FeH bands 
in cool, hydrogen-rich white dwarfs has not been included in the present work.

\subsection{Stellar parameters}

As a result of these recent studies, we have re-analysed the atmospheric 
properties of NLTT~7547 using two sets of hydrogen-rich model 
atmospheres. The first set consists of fully convective models and these are 
described in \citet{kaw2006} and \citet{kaw2012b}, however in these sets of 
models we have multiplied the self-broadening parameter, $\Gamma_{\rm BPO}$, 
from \citet{bar2000} by 1.0 rather than by 0.75 which was initially
adopted in the calculations of \citet{kaw2012b} in order to secure a mean sample mass
of $0.6\,M_\odot$. We then 
computed an additional set of model atmospheres and spectra where convection 
is suppressed and therefore energy is carried radiatively.

\begin{figure}
\includegraphics[width=0.9\columnwidth]{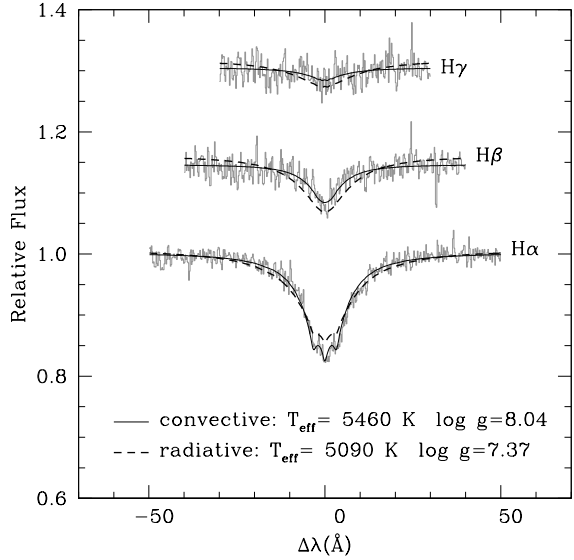}
\caption{Balmer line profiles showing Zeeman splitted cores of 
NLTT~7547 compared to the best fitting model spectrum with convection 
(solid lines) and convection fully suppressed (dashed lines).}
\label{fig_Balmer}
\end{figure}

Figure~\ref{fig_Balmer} shows the best-fitting synthetic Balmer lines compared 
to the X-shooter spectra for models including convection and, separately, for models that 
suppress convection. Using the convective models, we fitted the observed 
Balmer lines (H$\alpha$, H$\beta$ and H$\gamma$) of NLTT~7547 in the surface 
gravity ($\log{g}$) versus effective temperature ($T_{\rm eff}$) plane and 
obtained $T_{\rm eff} = 5460\pm80$ K and $\log{g} = 8.04\pm0.18$. Using the
non-convective models we obtained $T_{\rm eff} = 5090\pm100$ K and 
$\log{g} = 7.37\pm0.22$. However the calculated Balmer lines in the radiative 
models are poorly matched to the observed lines, particularly in the core. 
The superior goodness of fit of convective models relative to non-convective 
models ($\chi^2_{\rm R,nc}/\chi^2_{\rm R, c}=1.4$), which in principle is 
sufficient to reject non-convective models, may be an artifact of the 
line-broadening prescription. \citet{bed2017} analysed three magnetic
white dwarfs with effective temperature below 9000~K, and showed that in
these cases convective models were preferred and concluded that at lower 
effective temperatures it may be more difficult for a weak field to suppress
convection. However, the validity of our simple line broadening prescription
upon which our results are based remains to be confirmed in the presence 
of a $\approx 10^6$~G magnetic field.   

\begin{figure*}
\includegraphics[viewport=10 170 580 565,clip,width=0.7\textwidth]{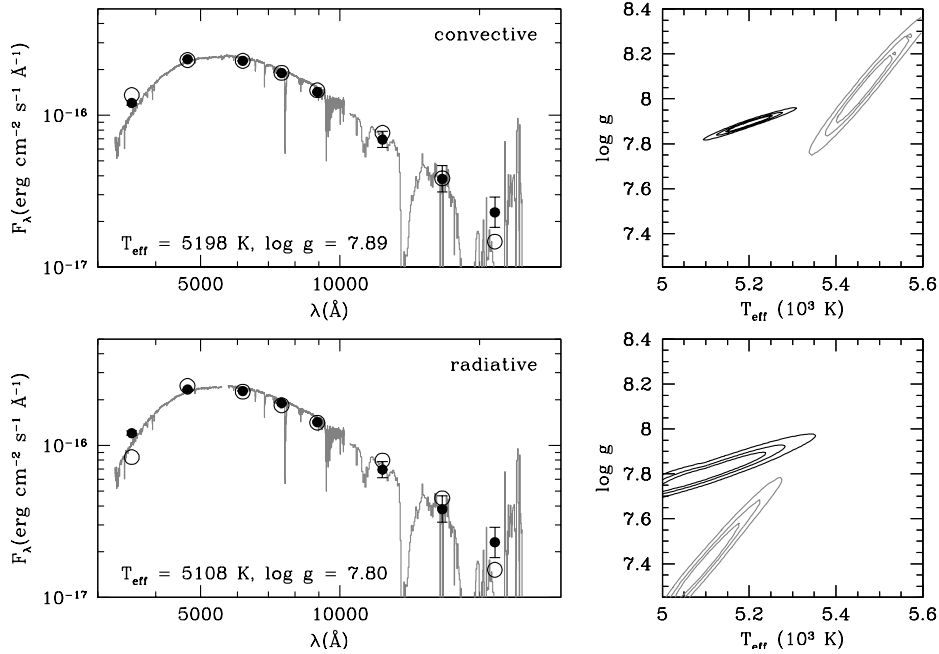}
\caption{The spectral energy distribution of NLTT~7547. The photometric 
measurements are represented by filled circles and the best-fitting model
photometric points are in open circles. The X-shooter spectrum (binned by 20 in 
UVB/VIS and 100 in NIR) is plotted in grey for comparison. The observed 
photometric points are compared to model photometry assuming a convective 
atmosphere (top panel) and assuming a radiative atmosphere (bottom panel). 
The right panels compare the 1,2 and 3-$\sigma$ contours of the SED fits 
(in black) and Balmer line fits (in grey).}
\label{fig_sed}
\end{figure*}

Next, we fitted the spectral energy distribution (SED) in the 
$T_{\rm eff}-\log{g}$ plane using available optical and infrared photometry. 
We constrained the synthetic magnitudes by using the Gaia parallax 
(Table~\ref{tab_phot}) to set the distance $D$ and by applying mass-radius 
relations from \cite{ben1999} to set the stellar radius $R$ in the ratio 
$R^2/D^2$. Figure~\ref{fig_sed} shows the observed photometric measurements 
from SDSS and 2MASS compared to the best fitting synthetic photometry. Using 
convective models we obtained $T_{\rm eff} = 5198\pm54$~K which is more than 
200~K lower than the effective temperature determined from the Balmer line 
fits. These results differ significantly and a similar disagreement between 
the effective temperature determined from Balmer lines and from the SED was 
encountered in the analysis of the cool DAZH NLTT~10480 \citep{kaw2011}.
Nearly identical parameters are obtained using radiative models but with a 
poorer goodness of fit. Figure~\ref{fig_sed} shows the observed SED compared 
to the best fitting synthetic photometry at $T_{\rm eff} = 5108\pm128$~K. 
After excluding the SDSS $u$ and 2MASS $K$ bands which deviate most from
the synthetic photometry we find the reduced $\chi^2$ for non-convective 
models $\chi^2_{R,nc} > 5$ while for convective models 
$\chi^2_{R,c} \approx 0.5$. The excess in the $K$ band may indicate the 
presence of circumstellar dust at a lower temperature than in any previous 
detections near white dwarfs \cite[see, ][]{fah2008}.

In summary, the Balmer lines and SED results are not mutually consistent 
using either convective or non-convective models. On the other hand 
synthetic Balmer line profiles and synthetic SED computed with convective 
models fit significantly better the data.    

We have calculated the Galactic velocity vectors of NLTT~7547 using the method
of \citet{joh1987} and the correction for the Solar motion from \citet{hog2005}.
As input we have used the distance and proper motion from Gaia and we measured
a barycentric corrected velocity of $\varv = -16.9$~km~s$^{-1}$ from the 
spectral lines in the X-shooter spectra. NLTT~7547 has components
$U, V, W = -44, -53, 52$~km~s$^{-1}$ which are representative of a thick disc 
velocity \citep{sou2003}. We have used \citet{ros2015} to calculate the 
z-component of the angular momentum ($J_z$) and the eccentricity ($e$) of
NLTT~7547. Following \citet{pau2003}, $J_z = 1420$~kpc~km~s$^{-1}$ and $e=0.262$
place NLTT~7547 between thin and thick disc Galactic orbits, making NLTT~7547
part of the old disc population.

Adopting the mass and radius constrained by the Gaia parallax in the SED
analysis (Table~\ref{tbl_prop}), the white dwarf cooling age is 3.1-4.3~Gyr.
The age of the thick disc has been estimated at 9.5-9.9~Gyr using white
dwarf luminosity function \citep{kil2017} or 10-13 Gyr based on cluster isochrones 
\citep[e.g., 47 Tuc, ][]{liu2000,sal2007}. The lifetime of the
progenitor would then extend between 6 to 10 Gyr corresponding to a mass
of 1.1 down to $0.95\,M_\odot$ using evolutionary models with a low metallicity 
\citep[$Z=0.004$, see a tabulation in][]{rom2015}
comparable to that of 47 Tuc. The initial mass (0.95-1.1$\,M_\odot$) and the 
corresponding final mass (0.50-0.54$\,M_\odot$) fit well within the
initial-to-final mass relations devised by, e.g., \citet{fer2005} and \citet{rom2015}.

\citet{tre2015} showed that the convection zone couples with the degenerate
core at an effective temperature of 5527~K for a $0.6\ M_\odot$ white dwarf
thereby increasing its energy release for a certain period of time. During that
period of time the cooling age of the non-magnetic (i.e., convective) white dwarf is 
effectively longer than that of its magnetic (i.e., with convection suppressed) counterpart
with the same effective temperature.
But having exhausted its internal energy faster, the non-magnetic white dwarf
catches up with its magnetic counterpart by the time they both reach a temperature
of 3340~K. 
At this stage the cooling rate of magnetic
white dwarfs decreases compared to the cooling rate of non-magnetic white dwarfs.
Below a temperature of 3340~K the roles are inverted and the cooling
age of the magnetic white dwarf becomes increasingly longer compared to a
non-magnetic white dwarf with the same effective temperature. 

At a lower mass of $0.52\ M_\odot$ appropriate for NLTT~7547, the coupling
would occur at a temperature lower than 5527~K and possibly as low as
as temperature of 5200~K appropriate for NLTT~7547. Therefore the cooling age
of convective models would overestimate the true age of NLTT~7547 assuming
that its magnetic field has suppressed the convection zone.
However, the difference in the cooling times 
amounts to $\lesssim$20\%, i.e., 0.7 Gyr.

\subsection{Magnetic fields}

\citet{kaw2012a} measured a longitudinal field of 
$B_l=-23.5\pm177.0$ kG using H$\beta$ and H$\gamma$ and $B_l=-32.2\pm40.2$ kG 
using the \ion{Ca}{ii} H\&K lines. The \ion{Ca}{ii} measurement has a 1-$\sigma$
upper limit of $\lvert B_l \rvert \lesssim \lvert 72.4\rvert$ kG. The large
uncertainty in the longitudinal magnetic field measurement does not allow us 
to place constraints on the inclination of the magnetic field.

The X-shooter spectra has revealed Zeeman splitted lines of Na, Mg, Al, Ca, Fe 
with H Balmer lines. Table~\ref{tbl_line} lists the heavy element lines 
detected and their rest wavelengths.

\begin{table}
\centering
\caption{Main spectral line identification} 
\label{tbl_line}
\begin{tabular}{lclc}
\hline
Ion & $\lambda$ (\AA) & Ion & $\lambda$ (\AA) \\
\hline
Fe{\sc i}  & 3719.934  & Ca{\sc ii} & 3968.469 \\
Ca{\sc ii} & 3736.902: & Ca{\sc i}  & 4226.728 \\
Fe{\sc i}  & 3737.131  & H$\beta$   & 4861.333 \\
Fe{\sc i}  & 3745.561  & Na {\sc i} & 5889.951 \\
Mg{\sc i}  & 3832.304  & Na {\sc i} & 5895.924 \\
Mg{\sc i}  & 3838.292  & H$\alpha$  & 6562.819 \\
Ca{\sc ii} & 3933.663  & Ca{\sc ii} & 8498.02: \\
Al{\sc i}  & 3944.006  & Ca{\sc ii} & 8542.09 \\
Al{\sc i}  & 3961.520  & Ca{\sc ii} & 8662.14 \\
\hline
\end{tabular}\\
\end{table}

The magnetic field of NLTT~7547 is low enough that the anomalous Zeeman 
splitting can be assumed for Na, Mg, Al, Ca and Fe. For details on how we 
calculate line splitting in this regime, see \citet{kaw2011}. We have
measured an average surface field of $155.7\pm12.4$ kG from the \ion{Ca}{ii} 
H line at 3933.663\AA\ while the Zeeman splitting in H$\alpha$ results in a 
surface average field of $163\pm4$ kG.

Depending on the topology of the magnetic field, the lines can be broadened as
a result of a field spread.
We have calculated model spectra by varying the dipolar field strength ($B_p$),
inclination ($i$) and the offset ($a_z$) by a fraction of the stellar radius
along the polar axis \citep[see][]{ven2018}. We have found that the best fit to
the data of NLTT~7547 is provided by a centred dipole ($a_z = 0$) of 240~kG.
A centred dipole field show broader lines due to field spread than a more
uniform field given by a substantially offset dipole. The broad line shapes of
metal lines in NLTT~7547 can be best modelled by a centred dipole ($a_z = 0$) 
of 240~kG in agreement with the H$\alpha$ modelling.

The centred dipole solution is not necessarily unique and
other combinations of field strength, inclination and offset are possible.
In order to investigate the likelihood of more complex field geometries, such
as the presence of a magnetic spot or multipolar fields, phase resolved data 
are required. Without any indication of phase-dependent variations we cannot
ascertain whether the star is in fact fast rotating and all measurements are
surface averaged, or the star is slow rotating and, therefore, is viewed at a
fixed phase angle. Consequently, the centred dipole model may not be complete, 
and we can only state that the broadened shape of the \ion{Ca}{ii} H\&K lines
is consistent with a centred dipole.

\cite{ven2018} revisited the cool, magnetic white dwarfs, NLTT~10480 \citep{kaw2011} 
and NLTT~53908 \citep{kaw2014} by modelling the magnetic field by considering
different field configurations. In our previous analysis of these objects, we 
assumed a uniform field. A dipolar field offset away from the observer is more
uniform and thus the spectral lines a less broadened. This seems to be the case
for NLTT~53908 which exhibits Zeeman features that are much sharper than those
yielded by a centred dipole. Our modelling indicates that the field structure
is that of a dipolar field of 635~kG offset by $a_z = -0.2$ (therefore away
from the observer) along the magnetic axis and $i=70^\circ$.
A refit of the narrow Zeeman Ca components in NLTT~10480 has also required 
a large offset of $a_z = -0.3$ for a field of $B_P=1.1$~MG and $i=70^\circ$.

\subsection{Abundance Analysis}

We determined the abundances of Na, Mg, Al, Ca and Fe using the effective
temperature and surface gravity derived by using our convective models. We calculated separately
the abundances for parameters obtained using Balmer line profiles and the SED.
The measured abundances are listed in Table~\ref{tbl_prop}.

We fitted all Ca lines, that is \ion{Ca}{ii} H\&K at 3933.663
and 3968.469 \AA, \ion{Ca}{i} at 4226.728 \AA\ and the infrared \ion{Ca}{ii} 
triplet at 8498.02, 8542.09 and 8662.14 \AA. We could not achieve a consistent
fit at the same abundance for all three sets of lines when using the atmospheric parameters from 
the Balmer line fits. Therefore, we fitted them separately and obtained
three separate abundance measurements. For \ion{Ca}{ii} H\&K we measured 
$\log{\rm (Ca/H)} = -10.06\pm0.06$, for \ion{Ca}{i} we obtained 
$\log{\rm (Ca/H)} = -9.60\pm0.18$ and using the \ion{Ca}{ii} infrared triplet 
we measured $\log{\rm (Ca/H)} = -8.91\pm0.23$. The consistency between the 
\ion{Ca}{ii} H\&K and \ion{Ca}{i} abundance measurements was improved by
decreasing the effective temperature, as was also shown by \citet{kaw2011} in 
the case of NLTT~10480. The discrepancy between \ion{Ca}{ii} H\&K and the 
infrared \ion{Ca}{ii} triplet remains unexplained.

Next, we repeated the calcium abundance measurements using the atmospheric 
parameters from the SED fit. Using 
\ion{Ca}{ii} H\&K we measured $\log{\rm (Ca/H)} = -10.12\pm0.04$, with
\ion{Ca}{i} we measured $\log{\rm (Ca/H)} = -9.77\pm0.36$, and with 
\ion{Ca}{ii} infrared triplet we measured $\log{\rm (Ca/H)} = -9.31\pm0.35$.
These abundance measurements are more consistent with each other than with the
warmer temperature derived from the Balmer line fits.

\begin{figure*}
\includegraphics[width=0.7\textwidth]{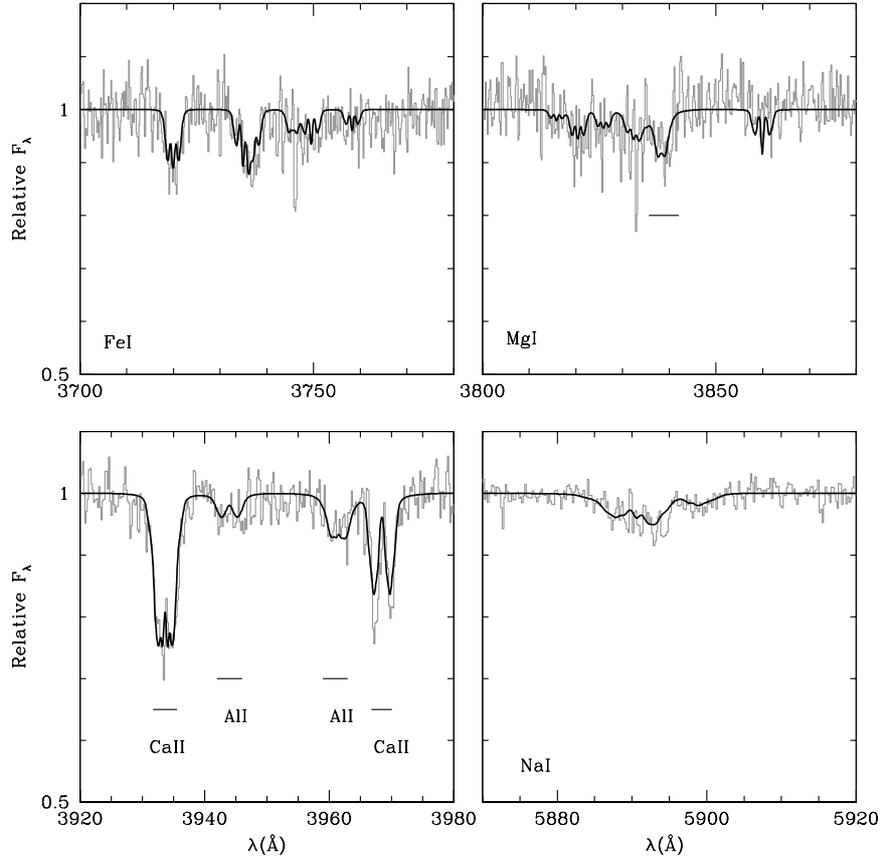}
\caption{X-Shooter spectrum (grey) compared to the best fitting model spectrum 
(black) at $T_{\rm eff} = 5460$~K and $\log{g} = 8.04$ showing the Zeeman 
splitted metal lines Na, Mg, Al, Ca and Fe.}
\label{fig_metals}
\end{figure*}

Figure~\ref{fig_metals} shows the X-shooter spectrum of NLTT~7547 compared to
the best-fitting model spectrum calculated using the atmospheric parameters 
($T_{\rm eff} = 5460$~K and $\log{g} = 8.04$) determined from the Balmer line fit. 
The main lines or group of lines labelled.

NLTT~7547 is yet another example of a
white dwarf accreting material from its circumstellar environment. Its 
composition, represented by Mg, Ca, and Fe, appears magnesium rich relative 
to bulk-Earth material \citep{hol2018}.

\begin{table}
\centering
\caption{Stellar and atmospheric parameters of NLTT~7547 using convective
models.}
\label{tbl_prop}
\begin{tabular}{lc}
\hline
Parameter & Measurement \\
\hline
\multicolumn{2}{c}{Balmer line profiles} \\
$T_{\rm eff}$ (K) & $5460\pm80$ \\
$\log{g}$ (cgs)   & $8.04\pm0.18$ \\
Mass (M$_\odot$ ) & $0.61\pm0.12$ \\
Distance (pc)     & $52.8\pm7.0$ \\
Age (Gyr)         & $3.5\pm1.0$ \\
$\log{\rm (Na/H)}$ & $-8.57\pm0.08$ \\
$\log{\rm (Mg/H)}$ & $-7.99\pm0.19$ \\
$\log{\rm (Al/H)}$ & $-8.79\pm0.20$ \\
$\log{\rm (Ca/H)}$ & $-10.06\pm0.06$ \\
$\log{\rm (Fe/H)}$ & $-8.81\pm0.13$ \\
\hline
\multicolumn{2}{c}{Spectral energy distribution} \\
$T_{\rm eff}$ (K) & $5198\pm54$ \\
$\log{g}$ (cgs)   & $7.89\pm0.04$ \\
Mass (M$_\odot$ ) & $0.52\pm0.02$ \\
Age (Gyr)         & $3.7\pm0.6$ \\
Radius ($R_\odot$)& $0.0135\pm0.0003$ \\
Gaia $\pi$ (mas)  & $22.55\pm0.16$ \\
Gaia Distance (pc)& $44.34\pm0.32$ \\
$\log{\rm (Na/H)}$ & $-8.68\pm0.09$ \\
$\log{\rm (Mg/H)}$ & $-8.09\pm0.14$ \\
$\log{\rm (Al/H)}$ & $-9.09\pm0.07$ \\
$\log{\rm (Ca/H)}$ & $-10.12\pm0.04$ \\
$\log{\rm (Fe/H)}$ & $-8.84\pm0.06$ \\
\hline
$B_S$ (kG)        & $163\pm4$ \\
\hline
\end{tabular}\\
\end{table}

\subsection{Test of variability}

Using H$\alpha$ we measured a surface averaged magnetic field 
$B_S = 163\pm4$ kG for NLTT~7547.
We also measured the surface averaged magnetic field using the 
Ca{\sc ii} H\&K lines. In this case we calculated model spectra at various
abundances and magnetic field strengths around 163 kG, and then by fitting
the observed Ca{\sc ii} H\&K line profiles we found $B_S = 155.7\pm10.5$ kG.
The dipolar field strength $B_P = 240$~kG is then constrained by the inclination
and dipole offset.
The resulting calcium abundance is $\log{\rm (Ca/H)}=-10.06\pm0.05$.
Since we obtained 6 sets of spectra over a period of less than one month, we
also fitted the Ca{\sc ii} H\&K line profiles for each observation to check for
variability. The measurements are summarized in Table~\ref{tbl_var}: the 
starting time and exposure time are for the UVB spectra and the Barrycentric
Julian Date (BJD) is calculated at mid-exposure. The Ca abundance and magnetic 
field strength do not appear to vary. We found that the average of Ca abundance 
is $\log{\rm (Ca/H)}=-10.05$ with a dispersion of 0.06 and the average of the 
magnetic field strength is $B_S = 155.4$ kG with a dispersion of 6.3 kG. Since 
the individual exposure times were 2940 sec each in the UVB arm, the white dwarf
could be either a fast rotator with a rotation period less than an hour or
a slow rotator with a period larger than a few months. Alternatively the field 
could be relatively stable as a function of the rotation phase, which would 
suggest very small viewing angles to the magnetic axis.

\begin{table*}
\centering
\caption{Calcium abundance and magnetic field measurements for individual spectroscopic
observations.}
\label{tbl_var}
\begin{tabular}{llcccc}
\hline
UT date & UT start & Exposure time (s) & BJD (2457200+) & $\log{\rm(Ca/H)}$ & $B_S$ (kG) \\
\hline
2015 August 24    & 09:09:35 & 2940 & 58.90153 & $-10.08\pm0.08$ & $158.1\pm16.5$ \\
2015 September 10 & 08:09:47 & 2940 & 75.86120 & $-10.12\pm0.11$ & $144.3\pm30.0$ \\
2015 September 13 & 08:28:18 & 2940 & 78.87424 & $-9.96\pm0.13$  & $161.5\pm20.5$ \\
2015 September 17 & 08:30:41 & 2940 & 82.87612 & $-10.11\pm0.09$ & $156.6\pm21.3$ \\
2015 September 19 & 05:05:32 & 2940 & 84.73375 & $-10.02\pm0.12$ & $150.0\pm22.7$ \\
2015 September 19 & 06:17:05 & 2940 & 84.78344 & $-10.02\pm0.11$ & $150.0\pm22.7$ \\
\hline
\end{tabular}
\end{table*}

We have used a fast Fourier transform routine to search for variations in the
photometric time series obtained in the $R$ filter, but did not yield any 
significant periods. Since this photometric time series was obtained during two
nights, we also
searched for periods in the data taken on both nights combined as well as data
obtained on individual nights. To place some constraints on the variations of 
NLTT~7547 we measured the dispersion of the photometric measurements.
On the first night (UT 18 October 2017) we measured a 
dispersion of $\sigma = 0.014$~mag and on the second night (UT 5 January 2018) 
we measured a dispersion of $\sigma = 0.011$~mag. The second night provided tighter
constraints as a result of better observing conditions. 

\section{Discussion}

The studies of cool and magnetic white dwarfs have unveiled new challenges
in our understanding of the atmospheric structure, formation and evolution
of magnetic white dwarfs. In this section we will discuss some of the issues
surrounding this class of objects.

\subsection{Convective versus radiative energy transport}

Magnetic fields have been thought to inhibit convection in cool white dwarfs. 
Some support for this suppression has recently come from the modelling of the 
weakly magnetic white dwarf WD~2105$-$820 \citep{gen2018} who fitted the 
ultraviolet and optical spectra of three non-magnetic white dwarfs with 
$T_{\rm eff}$ between 9000 and 11\,000~K and the weakly magnetic WD~2105$-$820 
with both radiative and convective models. For WD~2105$-$820, consistent 
atmospheric parameters were derived only when radiative models were used, 
while for the non-magnetic white dwarfs, consistent results were obtained with 
convective models. \citet{bed2017} provided independent evidence for convection 
suppression in WD~2105$-$820. Using radiative models, they were able to match 
the photometrically and spectroscopically determined effective temperatures and 
also successfully matched the spectroscopically determined distance with the 
measured parallax.

In the case of NLTT~7547, neither types of models successfully account for all 
data. Two independent diagnostics, the Balmer line profiles and the SED, 
provided irreconcilable stellar parameters ($T_{\rm eff},\ \log{g}$). However, 
convective models provided significantly better fits to the Balmer line 
profiles and the SED when analysed separately. Difficulties encountered in 
fitting Balmer line profiles may, in part, be caused by an inadequate line 
broadening theory in cool, magnetic atmospheres. We will assume that cool, 
magnetic white dwarfs have convective atmospheres
pending further investigations of non-convective atmospheres.

\subsection{Field structure}

\citet{bri2013}
obtained time-photometry series of several magnetic white dwarfs and found
that more than half are variable. Such white dwarfs can be used to study the 
magnetic field variations on the white dwarf surface as a function of
rotational phase. A few magnetic white dwarfs have been observed 
spectropolarimetrically over their rotational period revealing the complexity
of the magnetic field structure. The rapidly rotating EUVE~J0317-85.5 
varies photometrically, spectroscopically and spectropolarimetrically 
\citep{bar1995,fer1997,bur1999,ven2003}. These variations as a function of 
rotation phase reveal that the magnetic field structure of EUVE~J0317-85.5 is
rather complex and probably is a combination of a lower underlying field
of $\sim$185~MG with a stronger magnetic field of $\sim$450~MG concentrated 
in a spot \citep{ven2003}. More recently, the magnetic white dwarf in the
close double degenerate system NLTT~12758, showed photometric and 
spectropolarimetric variations as a function of the rotation phase 
\citep{kaw2017}. Photometric variations NLTT~8435 with a period of 95 minutes, 
that is mostly likely due to rotation, was reported by \citet{ven2018}.

Magnetic white dwarfs that do not appear to vary are either rotating with a 
very long period or their magnetic field is nearly aligned with the rotation 
axis. 
Photometric variations in NLTT~7547 are restricted to amplitudes below 11~mmag for
periods up to $\approx 2$ hr. 
In addition, we did not detect any variations in the six spectra obtained over several 
days, but because these spectra have exposure times close to one hour, any 
short period variations would be averaged out during the exposure and would remain
undetected. Higher signal-to-noise photometric time series would be required
to probe low-amplitude ($<10$ mmag), longer period ($>2$ hr) variations.

\subsection{Incidence of magnetism}

Recent identifications of magnetic white dwarfs suggest that the incidence
of magnetism differs among different spectral classes. In the past, various 
surveys have been carried out to search for new magnetic white dwarfs
to determine the incidence of magnetism. While colourimetric and magnitude
limited surveys \citep[e.g.,][]{sch1995,kep2013,bag2018a} yielded an incidence
of around 5\%, volume limited surveys delivered fractions as high as
20\% \citep[e.g.,][]{kaw2007}.

Most of these surveys were targeted toward hydrogen-rich white dwarfs. Recent 
observations of different spectral classes have revealed higher incidences
of magnetism in some of these classes. The highest fraction of magnetism
is found in the rare class of warm and hot DQ white dwarfs. 
\citet{duf2013} reported an incidence of $\sim 70$\%, which may suggest that
all warm/hot DQ may be magnetic. Hot DQ white dwarfs are photometrically
variable with periods ranging from a few minutes up to a couple of days. These
variations have been proposed to be the result of rotation 
\citep{law2013,wil2016}. \citet{duf2013} also suggested that this
group of white dwarfs are massive, and \cite{dun2015} proposed that they may
be the product of white dwarf mergers due to their rapid rotation and high mass
($M\gtrsim0.9\ M_\odot$).

\citet{hol2015,hol2017} have shown that the incidence of magnetism in a sample
of 79 cool 
($T_{\rm eff} < 8000$~K) DZ white dwarfs is more than 10\%, which is 
significantly higher than the incidence observed in the overall white dwarf 
population. The magnetic fields in these stars range from a $\approx 0.5$ MG
to over 10~MG. 

The sample of cool DAZ white dwarfs is still relatively small, but it already 
reveals that magnetism is more common among DAZs with $T_{\rm eff} < 6000$~K 
than in the general white dwarf population. 
\citet{kaw2014} have shown that over 40\% of DAZ white dwarfs with 
$T_{\rm eff} < 6000$~K are magnetic. 

The incidence of magnetism
among local ($\le 20$~pc) DA white dwarfs listed by \citet{gia2012} is
comparable ($\approx 20$\%) to that of all local white dwarfs \citep{kaw2007}.
However, the magnetic DA white dwarfs in the local sample cover a
much wider range of field strengths (0.1-100~MG) than that observed in the population
of magnetic DAZ white dwarfs (0.1-1~MG).
Therefore, the field incidence per decade of field strength is significantly larger
among DAZ white dwarfs. 
Note that taking into account the shorter cooling ages of
magnetic white dwarfs relative to their non-magnetic counterparts only
increases the significance of this excess.
On the other hand, the noted absence of high-field DAZ white dwarfs in our
sample underlines difficulties in detecting weak, heavily-spread spectral lines
in these objects.

\begin{figure}
\includegraphics[viewport=0 127 574 397,clip,width=\columnwidth]{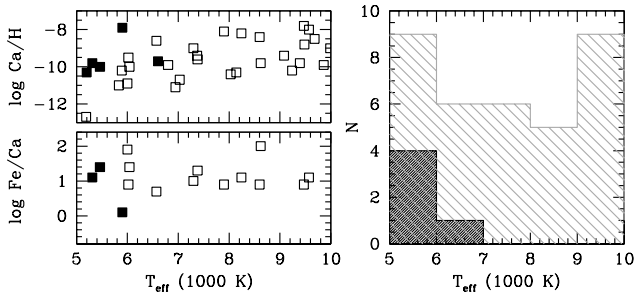}
\caption{{\it (Left:)} The abundances of cool DAZ white dwarfs showing 
$\log{\rm (Ca/H)}$ {\it (top)} and $\log{\rm (Fe/Ca)}$ {\it (bottom)}. The magnetic
white dwarfs are shown with full squares and non-magnetic with open squares. 
{\it (Right):} The temperature distribution of known DAZ white dwarfs with 
$T_{\rm eff} < 10\,000$ K are shown grey with the magnetic DAZ white dwarfs in
grey.}
\label{fig_abun_mag}
\end{figure}

Here we revisit this population study with an updated census which includes 
the case NLTT~7547 and others. Figure~\ref{fig_abun_mag} shows the known DAZ
white dwarfs that have been observed with sufficiently high resolution 
($R=\lambda/\Delta\lambda \gtrsim 5000$) that can constrain 
weak magnetic fields well below 100~kG,
while Table~\ref{tbl_cool} lists the properties of objects with an effective 
temperature below 7000~K. We also checked if these white dwarfs
were observed spectropolarimetrically, since spectropolarimetry can also reveal
weak magnetic fields if observed with high enough signal-to-noise ratio. We 
include longitudinal magnetic field measurements or upper limits for those stars
that have been observed spectropolarimetrically. 
Our sample has been selected from \citet{zuc2003}, \citet{far2011}, \citet{zuc2011}, 
\citet{kaw2011,kaw2012a,kaw2014,kaw2016}, \citet{kaw2011b} and \citet{ven2013}.
More cool DAZ white dwarfs have been discovered recently, however most of 
them have not been observed with sufficiently high resolution. Stars with 
$T_{\rm eff} < 7000$~K are WD2028$-$171 \citep{sub2017}, 
and WD0920$+$012 and WD1408$+$029 from \citet{say2012}. 
Another cool white dwarf, but with $T_{\rm eff} = 7220\pm246$~K
is WD2157$-$574 which was identified by \citet{sub2007}.
\citet{deb2010} reported the detection of \ion{Ca}{ii} H\&K in the nearby 
WD0141$-$675. We have analysed the X-shooter spectrum of WD0141$-$675 using our model
spectra to determine $T_{\rm eff} = 6150\pm10$~K, $\log{g} = 7.58\pm0.02$ and 
$\log{\rm (Ca/H)} = -10.96\pm0.11$. 
We have measured a velocity differential $\varv(\ion{Ca}{ii})-\varv({\rm H\alpha})\approx -3$~km s$^{-1}$ 
suggesting a photospheric origin to the calcium lines. 
The X-shooter spectrum does not
show evidence for the presence of a magnetic field. WD~1223$-$659 was reported 
to be a new DAZ by \citet{deb2010} with $T_{\rm eff} = 7660\pm220$~K 
\citep{sub2009}. The narrow Ca{\sc ii} line at 3933\AA\ shown in \citet{deb2010} 
does not appear to show signs of a magnetic field. 
WD~1223$-$659 is included in the histogram of 
Fig.~\ref{fig_abun_mag}, however since there are no abundance measurements
available it is excluded from the abundance plot. 
The figure shows that
the incidence of magnetism among the coolest known DAZ white dwarfs is now
closer to 50\% with a significantly lower incidence at higher
temperatures. 

\begin{table*}
\begin{minipage}{\textwidth}
\caption{Known cool DAZ white dwarfs ($T_{\rm eff} < 7000$ K) with mid/high-resolution spectra ($R \gtrsim 5000$). \label{tbl_cool}}
\centering
\begin{tabular}{@{}llccclccc@{}}
\hline
WD & Name & $T_{\rm eff}$ (K) & $\log{g}$ (c.g.s) & $\log{\rm (Ca/H)}$ & Fe$/$Ca & $B_s$ (kG) & $|B_l|$ (kG) &  Reference \\
\hline
WD0015$-$055 & NLTT~888 & $5570\pm40$  & $7.72\pm0.08$ & $-10.77\pm 0.06$ & 58    &  $<40$     & $<40$       &  1,2 \\
WD0028$+$220 & NLTT~1675 & $6020\pm50$  & $8.04\pm0.07$ & $-9.53\pm0.03$   &  8    &  $<40$     & ...         &  3 \\
WD0141$-$675 & LTT~934    & $6150\pm10$  & $7.58\pm0.02$ & $-10.96\pm0.11$  & ..    &  $<70$     & $<16$       &  4,5  \\
WD0151$-$308 & NLTT~6390 & $6040\pm40$  & $7.90\pm0.07$ & $-10.00\pm0.04$  & 27    &  $<40$     & $<11$       &  2,3 \\
WD0214$-$071 & NLTT~7547 & $5460\pm80$  & $8.04\pm0.18$ & $-10.01\pm0.05$  & 25    &  $163\pm4$ & $<72.4$     &  5 \\
WD0243$-$026 & LHS~1442   & $6800\pm300$ & $8.15\pm0.10$ & $-9.90$          & ...   &  $<10$     & ...         &  6 \\
WD0245$+$541 & G~174-14   & $5190\pm300$ & $8.22\pm0.10$ & $-12.69$         & ...   &  $<10$     & ...         &  7 \\
WD0315$-$293 & NLTT~10480 & $5200\pm200$ & $8.0\pm0.5$   & $-10.3\pm0.3$    & $<$10 &  $519\pm4$ & $212\pm50$  &  3,7 \\
WD0322$-$019 & G~77-50    & $5310\pm100$ & $8.05\pm0.01$ & $-9.8\pm0.2$     & 13    &  120       & $16.5\pm2.3$&  8,9 \\
WD0334$-$224 & NLTT~11393 & $5890\pm30$  & $7.86\pm0.06$ & $-10.24\pm0.04$  & $<$7  &  $<40$     & $<16$       &  2,3 \\
WD1208$+$576 & G~197-47   & $5830\pm300$ & $7.91\pm0.10$ & $-10.96$         & ...   &  $<10$     & ...         &  6 \\
WD1344$+$106 & G~63-54    & $6945\pm300$ & $7.99\pm0.10$ & $-11.13$         & ...   &  $<10$     & ...         &  6 \\
WD1633$+$433 & G~180-63   & $6570\pm300$ & $8.08\pm0.10$ & $-8.63$          & 5     &  $<10$     & ...         &  6 \\
WD1653$+$385 & NLTT~43806 & 5900         & 8.0           & $-7.9\pm0.19$    & 1.3   &   70       & ...         &  10 \\
WD2225$+$176 & NLTT~53908 & $6250\pm70$  & $7.87\pm0.12$ & $-9.85\pm 0.04$  & $<$13 &  $334\pm3$ & ...         &  1 \\
\hline
\end{tabular}\\
References: (1) \citet{kaw2014}; (2) \citet{kaw2012a}; (3) \citet{kaw2012b}; (4) \citet{kaw2007}; (5) This work; \\
(6) \citet{zuc2003}; (7) \citet{kaw2011}; (8) \citet{far2011}; (9) \citet{far2018}; (10) \citet{zuc2011}
\end{minipage}
\end{table*}

Recently, \citet{bri2018} discussed a possible scenario, originally 
proposed by \citet{far2011} to explain G~77-50, which may be able to explain
the higher incidence of magnetism among cool, magnetic DAZ and DZ white dwarfs.
In this scenario, a gaseous planet plunges onto an aging white dwarf. The 
differential rotation that is set off would generate a relatively weak magnetic 
field in the white dwarf. It is not clear what would cause such a planet to 
spiral in toward the white dwarf, but \citet{far2011} have proposed that outer 
planets and asteroids could have been destabilised by a close encounter with 
another stellar object during the billions of years that the white dwarf has
spent orbiting around the Galaxy. \citet{far2011} estimated that such an 
encounter has a 50 per cent probability to occur every 0.5 Gyr. Hence, the
older the white dwarf, the more likely it is that it will have experienced a 
close stellar encounter capable of perturbing the orbits of its outer planets 
that were not engulfed
during the progenitor star's giant phase of evolution. The accretion of rocky 
planets and asteroids pollutes white dwarfs, but should a giant
gaseous planet (a super-Jupiter) were to be present and accreted by
the white dwarf, a magnetic field could also be generated as a result and a
magnetic DAZ (or DZ) white dwarf would be born.

\section{Summary}

We have presented a spectroscopic analysis of NLTT~7547 showing that it is a 
new member of the population of cool and magnetic DAZ white dwarfs. With this 
addition we have shown that the incidence of magnetism is approximately 50\% 
for DAZ white dwarfs with $T_{\rm eff} < 6000$~K which decreases significantly 
to $\approx15\%$ for $6000 < T_{\rm eff} < 7000$~K and to a virtually zero 
incidence above $7000$~K. A single magnetic DAZ white dwarf, WD2105-820, is 
known to have a higher temperature \citep[$T_{\rm eff} = 10800$~K,][]{lan2012}. 
The higher incidence is, in part, accounted for by the increasing number of 
magnetic white dwarfs with age. Other mechanisms are required to explain the 
sharp increase in the incidence of magnetism in cool polluted white dwarfs. 
\cite{kaw2014} argued that the phenomenon coincides with the likely presence 
of a perennially crowded circumstellar environment which may have been caused
by the same event that generated the magnetic field.

The kinematics prescribed a membership to the old thin disc or
to the thick disc in agreement with a total age of 10 to 13 Gyr inferred from
a likely progenitor mass of 0.95 to 1.1\,$M_\odot$. Subtracting 0.7 Gyr to the total age of the star, as would
be mandated by shorter cooling ages of magnetic white dwarfs similar to NLTT~7547, does
not alter our conclusions concerning its population membership.

Challenges were encountered in the computation of cool, radiative atmospheres. 
The steep temperature gradient resulting from the suppression of convective 
energy transport is accompanied by a density reversal at the bottom of the 
atmosphere which could give rise to Rayleigh-Taylor instabilities. This 
eventuality was not explored further and future work should examine in detail 
the consequences of this phenomenon on the atmospheric structure. Remaining 
difficulties in reconciling stellar parameter measurements based on Balmer line 
profiles and the SED as well as inconsistent calcium abundance measurements 
require renewed modelling efforts aimed at resolving the competing effect of 
magnetic field and convective motion.

\section*{Acknowledgements}

AK and SV acknowledge support from the Czech Science Foundation
(15-15943S) and from the Mathematical Sciences Research Visitor 
Program of the Australian National University. EP acknowledges support from the
Ministry of Education of the Czech Republic grant LG15010 and Czech Science 
Foundation grant GA16-01116S. The International Centre for Radio Astronomy
Research is a joint venture between Curtin University and
the University of Western Australia, funded by the state
government of Western Australia and the joint venture
partners. This work has made use of 
the VALD database, operated at Uppsala University, the Institute of Astronomy 
RAS in Moscow, and the University of Vienna. This study is based on 
observations made with ESO telescopes at the La Silla Paranal Observatory 
under programmes 095.D-0311 and 099.D-0661. This is based in part on data 
collected with the Danish 1.54-m telescope at the ESO La Silla Observatory.
We thank Dayal Wickramasinghe for stimulating discussions. We thank the 
referee Dr. Stefano Bagnulo for useful comments.

\bsp	
\label{lastpage}

\begin{thebibliography}{99}
\bibitem[\protect\citeauthoryear{Achilleos \& Wickramasinghe}{1989}]{ach1989} 
Achilleos N., Wickramasinghe D.~T., 1989, ApJ, 346, 444
\bibitem[\protect\citeauthoryear{Albareti et al.}{2017}]{alb2017} 
Albareti F.~D. et al., 2017, ApJS, 233, 25
\bibitem[\protect\citeauthoryear{Bagnulo et al.}{2018}]{bag2018a} Bagnulo S.,
Landstreet J.~D., Martin A.~J., Valyavin G., 2018, Contrib. Astron. Obs. 
Skalnate Pleso, 48, 236
\bibitem[\protect\citeauthoryear{Bagnulo \& Landstreet}{2018}]{bag2018b} 
Bagnulo S., Landstreet J.~D., 2018, A\&A, 618, A113
\bibitem[\protect\citeauthoryear{Barklem \& Collet}{2016}]{bar2016} 
Barklem P.~S., Collet R., 2016, A\&A, 588, A96
\bibitem[\protect\citeauthoryear{Barklem et al.}{2000}]{bar2000} Barklem P.~S., 
Piskunov N., O'Mara B.~J., 2000, A\&A, 363, 1091
\bibitem[\protect\citeauthoryear{Barstow, et al.}{1995}]{bar1995} Barstow M.~A.,
Jordan S., O'Donoghue D., Burleigh M.~R., Napiwotzki R., Harrop-Allin M.~K., 
1995, MNRAS, 277, 971
\bibitem[\protect\citeauthoryear{B{\'e}dard et al.}{2017}]{bed2017} B{\'e}dard A.,
Bergeron P., Fontaine G., 2017, ApJ, 848, 11
\bibitem[\protect\citeauthoryear{Benvenuto \& Althaus}{1999}]{ben1999} 
Benvenuto O.~G., Althaus L.~G., 1999, MNRAS, 303, 30
\bibitem[\protect\citeauthoryear{Braithwaite \& Spruit}{2004}]{bra2004}
Braithwaite J., Spruit H.~C. 2004, Nature, 431, 819
\bibitem[\protect\citeauthoryear{Briggs et al.}{2018}]{bri2018} Briggs G.~P., 
Ferrario L., Tout C.~A., Wickramasinghe D.~T., 2018, MNRAS, 478, 899
\bibitem[\protect\citeauthoryear{Brinkworth et al.}{2013}]{bri2013} 
Brinkworth C.~S., Burleigh M.~R., Lawrie K., Marsh T.~R., Knigge C., 2013,
ApJ, 773, 47
\bibitem[\protect\citeauthoryear{Brown et al.}{2018}]{bro2018} 
Brown A.~G.~A. et al., 2018, A\&A, 616, A1
\bibitem[\protect\citeauthoryear{Burleigh et al.}{1999}]{bur1999} 
Burleigh M.~R., Jordan S., Schweizer W., 1999, ApJ, 510, L37
\bibitem[\protect\citeauthoryear{Cheng et al.}{2017}]{che2017} Cheng L.,
Gauss J., Ruscic, B., Armentrout P.~B., Stanton J.~F., 2017, 
J. Chem. Theory Comput., 13, 1044
\bibitem[\protect\citeauthoryear{D'Antona \& Mazzitelli}{1975}]{dan1975}
D'Antona F., Mazzitelli I., 1975, A\&A, 42, 127
\bibitem[\protect\citeauthoryear{Debes \& Kilic}{2010}]{deb2010} Debes J.~H., 
Kilic M. 2010, American Institute of Physics Conference Series, 1273, 488 
\bibitem[\protect\citeauthoryear{Debes \& Sigurdsson}{2002}]{deb2002} 
Debes J.~H., Sigurdsson S., 2002, ApJ, 572, 556 
\bibitem[\protect\citeauthoryear{Dufour et al.}{2013}]{duf2013} Dufour P.,
Vornanen T., Bergeron P., Fontaine G., Berdyugin A., 2013, 18th European 
Workshop on White Dwarfs, ASP Conf. Ser., Vol. 469, eds. J. Krzesi{\'n}ski, 
G. Stachowski, P. Moskalik, K. Bajan, 167
\bibitem[\protect\citeauthoryear{Dunlap \& Clemens}{2015}]{dun2015} 
Dunlap B.~H., Clemens J.~C., 2015, in ASP Conf. Ser., Vol. 493, eds. P. Dufour,
P. Bergeron, G. Fontaine, 547
\bibitem[\protect\citeauthoryear{Euchner et al.}{2005}]{euc2005} Euchner F.,
Reinsch K., Jordan S., Beuermann K., G{\"a}nsicke B.~T., 2005, A\&A, 442, 651
\bibitem[\protect\citeauthoryear{Farihi et al.}{2011}]{far2011} Farihi J., 
Dufour P., Napiwotzki R., Koester D. 2011, MNRAS, 413, 2559
\bibitem[\protect\citeauthoryear{Farihi et al.}{2018}]{far2018} Farihi J., 
et al., 2018, MNRAS, 474, 947
\bibitem[\protect\citeauthoryear{Farihi, Zuckerman \& Becklin}{2008}]{fah2008} 
Farihi J., Zuckerman B., Becklin E.~E., 2008, ApJ, 674, 431
\bibitem[\protect\citeauthoryear{Ferrario et al.}{2015}]{fer2015} Ferrario L.,
de Martino D., G\"{a}nsicke B.~T., 2015, Space Sci. Rev., 191, 111
\bibitem[\protect\citeauthoryear{Ferrario et al.}{1997}]{fer1997} Ferrario L., 
Vennes S., Wickramasinghe D.~T., Bailey J.~A., Christian D.~J., 1997, MNRAS, 
292, 205
\bibitem[\protect\citeauthoryear{Ferrario et al.}{2005}]{fer2005} Ferrario L., 
Wickramasinghe D., Liebert J., Williams K.~A., 2005, MNRAS, 361, 1131
\bibitem[\protect\citeauthoryear{Garc{\'\i}a-Berro et al.}{2012}]{gar2012} 
Garc{\'\i}a-Berro E., et al., 2012, ApJ, 749, 25
\bibitem[\protect\citeauthoryear{Gentile Fusillo et al.}{2018}]{gen2018}
Gentile Fusillo N.~P., Tremblay P.-E., Jordan S., G{\"a}nsicke B.~T., 
Kalirai J.~S., Cummings J., 2018, MNRAS, 473, 3693
\bibitem[\protect\citeauthoryear{Giammichele et al.}{2012}]{gia2012} 
Giammichele N., Bergeron P., Dufour P., 2012, ApJS, 199, 29
\bibitem[\protect\citeauthoryear{Hogg et al.}{2005}]{hog2005} Hogg D.~W., 
Blanton M.~R., Roweis S.~T., Johnston K.~V., 2005, ApJ, 629, 268
\bibitem[\protect\citeauthoryear{Hollands et al.}{2015}]{hol2015} 
Hollands M.~A., G{\"a}nsicke B.~T., Koester D., 2015, MNRAS, 450, 681
\bibitem[\protect\citeauthoryear{Hollands, G{\"a}nsicke \& Koester}{2018}]{hol2018} 
Hollands M.~A., G{\"a}nsicke B.~T., Koester D., 2018, MNRAS, 477, 93
\bibitem[\protect\citeauthoryear{Hollands et al.}{2017}]{hol2017} 
Hollands M.~A., Koester D., Alekseev V., Herbert E.~L., G{\"a}nsicke B.~T. 2017,
MNRAS, 467, 4970
\bibitem[\protect\citeauthoryear{Johnson \& Soderblom}{1987}]{joh1987} 
Johnson D.~R.~H., Soderblom D.~R., 1987, AJ, 93, 864
\bibitem[\protect\citeauthoryear{Jura}{2008}]{jur2008} Jura M., 2008, AJ, 135, 1785 
\bibitem[\protect\citeauthoryear{Jura et al.}{2007}]{jur2007} Jura M., 
Farihi J., Zuckerman B., 2007, ApJ, 663, 1285
\bibitem[\protect\citeauthoryear{Kawka et al.}{2017}]{kaw2017} Kawka A., 
Briggs G.~P., Vennes S., Ferrario L., Paunzen E., Wickramasinghe D.~T., 2017, 
MNRAS, 466, 1127
\bibitem[\protect\citeauthoryear{Kawka \& Vennes}{2006}]{kaw2006} Kawka A.,
Vennes S., 2006, ApJ, 643, 402
\bibitem[\protect\citeauthoryear{Kawka \& Vennes}{2011}]{kaw2011} Kawka A.,
Vennes S., 2011, A\&A, 532, A7
\bibitem[\protect\citeauthoryear{Kawka et al.}{2011}]{kaw2011b} Kawka A.,
Vennes S., Dinnbier F., Cibulkov{\'a} H., N{\'e}meth P. 2011, American 
Institute of Physics Conference Series, 1331, 238
\bibitem[\protect\citeauthoryear{Kawka \& Vennes}{2012a}]{kaw2012a} Kawka A.,
Vennes S., 2012a, MNRAS, 425, 1394
\bibitem[\protect\citeauthoryear{Kawka \& Vennes}{2012b}]{kaw2012b} Kawka A.,
Vennes S., 2012b, A\&A, 538, A13
\bibitem[\protect\citeauthoryear{Kawka \& Vennes}{2014}]{kaw2014} Kawka A.,
Vennes S., 2014, MNRAS, 439, L90
\bibitem[\protect\citeauthoryear{Kawka \& Vennes}{2016}]{kaw2016} Kawka A.,
Vennes S., 2016, MNRAS, 458, 325
\bibitem[\protect\citeauthoryear{Kawka et al.}{2007}]{kaw2007} Kawka A.,
Vennes S., Schmidt G.D., Wickramasinghe D.T., Koch R., 2007, ApJ, 654, 499 
\bibitem[\protect\citeauthoryear{Kemic}{1975}]{kem1975} Kemic S.~B., 1975, 
Ap\&SS, 36, 459
\bibitem[\protect\citeauthoryear{Kepler et al.}{2013}]{kep2013} Kepler S.~O.
et al., 2013, MNRAS, 429, 2934
\bibitem[\protect\citeauthoryear{Kilic et al.}{2006}]{kil2006} Kilic M., 
von Hippel T., Leggett S.~K., Winget D.~E., 2006, ApJ, 646, 474
\bibitem[\protect\citeauthoryear{Kilic et al.}{2017}]{kil2017} Kilic M., 
et al., 2017, ApJ, 837, 162
\bibitem[\protect\citeauthoryear{Landi Degl'Innocenti \& Landolfi}{2004}]{lan2004} 
Landi Degl'Innocenti E., Landolfi M., 2004, Polarisation in spectral lines, No. 307 
in Astrophysics and Space Library, Kluwer Academic Publishers, Dordrecht ASSL, 307
\bibitem[\protect\citeauthoryear{Landstreet et al.}{2012}]{lan2012}
Landstreet J.~D., Bagnulo S., Valyavin G.~G., Fossati L., Jordan S., Monin D., 
Wade G.~A., 2012, A\&A, 545, A30
\bibitem[\protect\citeauthoryear{Landstreet et al.}{2017}]{lan2017} 
Landstreet J.~D., Bagnulo S., Valyavin G.~G., Valeev A.~F., 2017, A\&A, 607,
A92
\bibitem[\protect\citeauthoryear{Lawrie et al.}{2013}]{law2013} Lawrie K.~A., 
Burleigh M.~R., Dufour P., Hodgkin S.~T., 2013, MNRAS, 433, 1599
\bibitem[\protect\citeauthoryear{Liebert \& Sion}{1979}]{lie1979} Liebert J., 
Sion, E.~M., 1979, ApJ, 20, L53
\bibitem[\protect\citeauthoryear{Liu \& Chaboyer}{2000}]{liu2000} Liu W.~M., 
Chaboyer B., 2000, ApJ, 544, 818
\bibitem[\protect\citeauthoryear{Martin \& Wickramasinghe}{1984}]{mar1984} 
Martin B., Wickramasinghe D.~T., 1984, MNRAS, 206, 407 
\bibitem[\protect\citeauthoryear{Maxted et al.}{2000}]{max2000} Maxted P.~F.~L.,
Ferrario, L., Marsh T.~R., Wickramasinghe D.~T., 2000, MNRAS, 315, L41
\bibitem[\protect\citeauthoryear{Pauli et al.}{2003}]{pau2003} Pauli E.-M., 
Napiwotzki R., Altmann M., Heber U., Odenkirchen M., Kerber F., 2003, A\&A, 
400, 877
\bibitem[\protect\citeauthoryear{Romero et al.}{2015}]{rom2015} Romero A.~D.,
Campos F., Kepler S.~O., 2015, MNRAS, 450, 3708
\bibitem[\protect\citeauthoryear{Rossi}{2015}]{ros2015} Rossi L.~J., 2015, 
A\&C, 12, 11
\bibitem[\protect\citeauthoryear{Salaris et al.}{2007}]{sal2007} Salaris M., 
Held E.~V., Ortolani S., Gullieuszik M., Momany Y., 2007, A\&A, 476, 243
\bibitem[\protect\citeauthoryear{Salim \& Gould}{2003}]{sal2003} Salim S.,
Gould A., 2003, ApJ, 582, 1011
\bibitem[\protect\citeauthoryear{Sauval \& Tatum}{1984}]{sau1984} Sauval A.~J.,
Tatum J.~B., 1984, ApJS, 56, 193
\bibitem[\protect\citeauthoryear{Sayres et al.}{2012}]{say2012} Sayres C., 
Subasavage J.~P., Bergeron P., Dufour P., Davenport J.~R.~A., AlSayyad Y., 
Tofflemire B.~M., 2012, AJ, 143, 103
\bibitem[\protect\citeauthoryear{Schimeczek \& Wunner}{2014}]{sch2014} 
Schimeczek C., Wunner G., 2014, ApJS, 212, 26
\bibitem[\protect\citeauthoryear{Schmidt \& Smith}{1995}]{sch1995} 
Schmidt G.~D., Smith P.~S., 1995, ApJ, 448, 305
\bibitem[\protect\citeauthoryear{Skrutskie et al.}{2006}]{skr2006}
Skrutskie M.~F. et al., 2006, AJ, 131, 1163
\bibitem[\protect\citeauthoryear{Soubiran et al.}{2003}]{sou2003}
Soubiran C., Bienaym\'e O., Siebert A., 2003, A\&A, 398, 141
\bibitem[\protect\citeauthoryear{Subasavage et al.}{2007}]{sub2007} 
Subasavage J.~P., Henry T.~J., Bergeron P., Dufour P., Hambly N.~C., 
Beaulieu T.~D., 2007, AJ, 134, 252
\bibitem[\protect\citeauthoryear{Subasavage et al.}{2009}]{sub2009} 
Subasavage J.~P., Jao W.-C., Henry T.~J., Bergeron P., Dufour P., Ianna P.~A., 
Costa E., M{\'e}ndez R.~A., 2009, AJ, 137, 4547
\bibitem[\protect\citeauthoryear{Subasavage et al.}{2017}]{sub2017} 
Subasavage J.~P., et al., 2017, AJ, 154, 32
\bibitem[\protect\citeauthoryear{Tremblay et al.}{2015}]{tre2015} 
Tremblay P.-E., Fontaine G., Freytag B., Steiner O., Ludwig H.-G., Steffen M.,
Wedemeyer S., Brassard P., 2015, ApJ, 812, 19 
\bibitem[\protect\citeauthoryear{Valyavin et al.}{2008}]{val2008} Valyavin G.,
Wade G.~A., Bagnulo S., Szeifert T., Landstreet J.~D., Han I., Burenkov A., 
2008, ApJ, 683, 466
\bibitem[\protect\citeauthoryear{Valyavin et al.}{2014}]{val2014} Valyavin G.,
et al., 2014, Nature, 515, 88
\bibitem[\protect\citeauthoryear{Valyavin \& Fabrika}{1999}]{val1999} 
Valyavin, G., Fabrika S., 1999, 11th European Workshop on White Dwarfs, ASP 
Conf. Ser., Vol. 169, eds. S.-E. Solheim, E.G. Meistas, 206
\bibitem[\protect\citeauthoryear{Vennes \& Kawka}{2013}]{ven2013} Vennes S.,
Kawka A., 2013, ApJ, 779, 70
\bibitem[\protect\citeauthoryear{Vennes et al.}{2018}]{ven2018} Vennes S., 
Kawka A., Ferrario L., Paunzen E., 2018, CoSka, 48, 307
\bibitem[\protect\citeauthoryear{Vennes et al.}{2003}]{ven2003} Vennes S.,
Schmidt G.~D., Ferrario L., Christian D.~J., Wickramasinghe D.~T., Kawka A.,
2003, ApJ, 593, 1040
\bibitem[\protect\citeauthoryear{Vernet et al.}{2011}]{ver2011} Vernet J. 
et al., 2011, A\&A, 536, A105
\bibitem[\protect\citeauthoryear{Wende et al.}{2010}]{wen2010} Wende S., 
Reiners A., Seifahrt A., Bernath P.~F., 2010, A\&A, 523, A58
\bibitem[\protect\citeauthoryear{Wickramasinghe \& Martin}{1986}]{wic1986}
Wickramasinghe D.~T., Martin B., 1986, MNRAS, 223, 323
\bibitem[\protect\citeauthoryear{Williams et al.}{2016}]{wil2016} 
Williams K.~A., Montgomery M.~H., Winget D.~E., Falcon R.~E., Bierwagen M., 
2016, ApJ, 817, 27
\bibitem[\protect\citeauthoryear{Zuckerman et al.}{2003}]{zuc2003} 
Zuckerman B., Koester D., Reid I.~N., H{\"u}nsch M. 2003, ApJ, 596, 477
\bibitem[\protect\citeauthoryear{Zuckerman et al.}{2010}]{zuc2010}
Zuckerman B., Melis C., Klein B., Koester D., Jura M., 2010, ApJ, 722, 725
\bibitem[\protect\citeauthoryear{Zuckerman et al.}{2011}]{zuc2011} 
Zuckerman B., Koester D., Dufour P., Melis C., Klein B., Jura M. 2011, 
ApJ, 739, 101
\end{thebibliography}
\end{document}